\documentclass[10pt,preprint]{aastex}
\usepackage{amsmath}
\usepackage{graphicx}
\usepackage{epsfig}
\usepackage[title]{appendix}
\usepackage{hyperref}
\usepackage{natbib}
\usepackage{fancyvrb}
\DefineVerbatimEnvironment{code}{Verbatim}{fontsize=\small}
\DefineVerbatimEnvironment{example}{Verbatim}{fontsize=\small}
\usepackage{listings}
\usepackage{ulem,color}

\title{\MakeUppercase{The EB Factory Project I. A Fast, Neural Net Based, General Purpose Light Curve Classifier Optimized for Eclipsing Binaries}}
\author{Martin Paegert$^{1}$, Keivan G. Stassun$^{1, 2}$, Dan M. Burger$^{1}$}
\affil{$^{1}$Vanderbilt University, Nashville, TN 37235, USA}
\affil{$^{2}$Fisk University, Nashville, TN 37208, USA}

\begin{document}



\begin{abstract}
We describe a new neural-net based light curve classifier and provide it with documentation as a ready-to-use tool for the community. While optimized for identification and classification of eclipsing binary stars, the classifier is general purpose, and has been developed for speed in the context of upcoming massive surveys such as LSST. A challenge for classifiers in the context of neural-net training and massive data sets is to minimize the number of parameters required to describe each light curve. We show that a simple and fast geometric representation that encodes the overall light curve shape, together with a chi-square parameter to capture higher-order morphology information results in efficient yet robust light curve classification, especially for eclipsing binaries. Testing the classifier on the ASAS light curve database, we achieve a retrieval rate of 98\% and a false-positive rate of 2\% for eclipsing binaries.  We achieve similarly high retrieval rates for most other periodic variable-star classes, including RR Lyrae, Mira, and delta Scuti. However, the classifier currently has difficulty discriminating between different sub-classes of eclipsing binaries, and suffers a relatively low ($\sim$60\%) retrieval rate for multi-mode delta Cepheid stars. We find that it is imperative to train the classifier's neural network with exemplars that include the full range of light curve quality to which the classifier will be expected to perform; the classifier performs well on noisy light curves only when trained with noisy exemplars. The classifier source code, ancillary programs, a trained neural net, and a guide for use, are provided.
\end{abstract}

\textit{Accepted by AJ, \date{May 2014}}


\section{Introduction}

Eclipsing binary (EB) stars are of high importance for determining calibration-free stellar properties like masses, surface temperatures, radii and luminosities (see \citet{torres10}), stellar distances (\citet{guinan98}), ages (\citet{thompson10}), and for testing stellar evolutionary models (\citet{stassun09}). The results of modeling EBs are used in many areas of astrophysics, including calibrating cosmic distances, mapping clusters and other stellar populations in the Milky Way, energy transfer mechanisms and calibrating fundamental relations as mass-radius-luminosity and color-temperature relationship. Current and future ground-based and space-based surveys such as \textit{Kepler} \citep{kepler10}, the Large Synoptic Survey Telescope (LSST) \citep[see][]{prsa11} and the Transiting Exoplanet Survey Satellite \citep{tess} produce many thousands to millions of light curves at a time, many of which are EBs.

Identifying EBs in survey data has until now been mainly a manual effort. In the first release of the \textit{Kepler} Binary catalog \citet{keplerebs1} state: "preliminary classification was done manually by visual target inspection". EBs continue to be carefully, but manually vetted by the \textit{Kepler} EB science team \citep{keplerebs} and so far about $2200$ EBs have been found in the $156,000$ stars observed by \textit{Kepler}. This yields 1.4\% which is roughly as expected from simple geometry. Following the estimations in \citet{prsa11}, LSST will observe about 2 billion stars yielding 28 million EBs. Due to the less than optimal coverage in time 28\% or 7.8 million of these will be detectable, a factor of several thousand increase over the \textit{Kepler} EB Catalog. Clearly these numbers indicate that a manual approach to light curve classification for analysis of EBs cannot continue into the LSST era.

The EB Factory \citep{ebfabs} is an end-to-end computational pipeline that allows automatic processing of massive amounts of light curve data --- from period finding, to object classification, to determination of the stellar physical properties --- in order to find the most scientifically interesting EBs and to permit accurate modeling of these EBs for detailed tests and benchmarking of theoretical stellar evolution models. The EB Factory processes survey light curve data in the following 'production steps', each of which is being developed as an independent algorithmic module: (1) Period search and phase folding, (2) Classification, (3) EB Solution Estimator and (4) EB Solution Refiner. The first  step determines whether the light curve exhibits periodic behavior. Periodic objects determined in the classification step to be high likelihood EBs are then passed to a neural net or genetic algorithm based solution estimator \citep[see][]{prsa08} that estimates the underlying physical properties of the EBs (i.e. stellar dimensions, temperatures, etc). Those deemed most scientifically interesting based on these estimated properties are then subjected to detailed analysis by the Solution Refiner (currently based on the PHOEBE EB modeling engine; \citet{prsa08}).

The focus of this paper is the light curve classification step. The classification step is important, because it distinguishes the basic astrophysical nature of different kind of periodic variables, and allows one to invoke the most appropriate specific analysis procedures for the subsequent analysis steps. Many different types of stars exhibit periodic variability aside from the EBs which are the main subject of our effort (e.g., Cepheids, RR Lyrae, delta Scuti stars). We seek to make the light curve classification step likelihood-based so that the user can decide appropriate confidence thresholds, depending on the user's needs for sample completeness versus fidelity. In other words, we seek a classification approach that can output a quantitative probability that a given light curve belongs to a specific class of physical object. 

Because different classes of variable stars show distinctive patterns in their light curves, and because neural networks have their strength in pattern recognition, classification of light curves via neural networks has been investigated by several authors. \citet{eyer05} used unsupervised learning to identify statistical clusters in parameter space which then were associated to main variable classes such as EBs, RR Lyrae, Cepheids and others. \citet{debosscher07} analyzed different machine learning techniques, including neural networks, which have then been tested by \citet{sarro09} using the Optical Garvitational Lensing Experiment (OGLE-II) catalog \citep{ogleii,oglemaps} and applied to light curves from \textit{CoRoT} \citep{debosscher09}. An improved version has been used with the TrES Lyr1 field; \citet{blomme11} reports retrieval rates between 87\% and 98\% for EBs aggregated across EB sub-classes. Recently \citet{richards12} and \citet{hippcls} applied an alternative technique (random forests) to a re-classified All Sky Automatted Survey (ASAS) catalog and to variables in the \textit{Hipparcos} catalog.

However, a crucial challenge with neural networks, especially in the context of very large data sets, is to optimize the manner in which the light curve data are represented. In order to permit rapid processing of large data sets, the form used to represent each light curve should itself be computationally fast, and the number of parameters required to represent each light curve should be small. This must be balanced with ensuring that the representational form has sufficiently high fidelity to capture the features needed for accurate classification. Inclusion of too few parameters or of insignificant features can lead to sub-optimal classification.

A confounding feature of astronomical light curve data in particular is that in general they are not homogenous in time. When processing time-series data, neural networks generally require equidistant measurements in time, much like standard Fourier analysis methods for period finding. Aside from the diurnal and seasonal cycles, weather or technical problems can cause gaps in observational data. Phase-folding a light curve around its main period helps in terms of creating a denser representation of the light curve. But even a phase-folded light curve may show gaps, rarely having measurements equidistant in phase. 

One common way to represent the light curve data is to pre-process each light curve via a Fourier analysis and to characterize the light curve by the strongest frequencies or by a function of these frequencies. This standard technique is slow, because Fourier algorithms are inherently computationally expensive, and in order to fully represent the data typically leads to 30 or more input parameters resulting in overall slower classifications (a more detailed description can be found in \citet{debosscher07}). Moreover, neural networks need a training set of classified light curves to learn from and the number of such training light curves needed scales approximately quadratic with the number of input parameters. In addition, rare classes --- classes with few examples in the training set --- tend to be under-trained and thus show low retrieval rates; so reducing the number of input parameters allows a more fine-grained classification and thus improves training and recovery of rare classes. In addition, for EBs, representing a light curve by a Fourier series (as in \citet{asasebs} and \citet{debosscher07}) encounters problems for especially wide binaries with narrow, deep eclipses, because they possess many high-order Fourier coefficients and thus increases the number of input parameters required to faithfully represent the data.

In this paper we introduce a simple, yet fast and robust method of light curve classification, optimized for EBs (including those with narrow, deep eclipses). We represent each phase-folded light curve with a chain of four second-order polynomials, and use the polynomial coefficients and positions as the primary input parameters to the classifier. We also include the period, light curve amplitude, and the $\chi^2$ of the polynomial fit as parameters. Thus the classification is performed with only 19 parameters characterizing each light curve. This represents a reduction of at least $1/3$ in the number of required input parameters compared to Fourier-based light curve classification methods discussed above, and the parameters are moreover easy and fast to compute, making the overall classification process very fast. We test our approach with the ASAS catalog \citep{asas}, consisting of 46744 classified light curves for 13 different classes of variable stars. Thus the method is general and can be applied to all types of variable stars. For EBs the success rate is shown to be up to 98.3\%.

In Section \ref{selection} we describe the data that we use for training and testing of our light curve classifier. In Section \ref{sneuralnet} we present our light curve classifier, including the neural network setup and our method for characterizing each light curve with a minimal set of input parameters. Section \ref{results} presents the results of applying our classifier to the ASAS data set and a first glimpse on running a trained network over \textit{Kepler} light curves. We discuss our results in Section \ref{discussion} and summarize in Section \ref{summary}. Finally, in an Appendix we document our implementation for use by the community, and the source code together with a ready-to-use trained neural network is available for download. \footnote{http://www.vanderbilt.edu/AnS/physics/vida/ebfactory.htm}

\section{\label{selection} Light Curve Data for Classifier Training and Testing}

In order to train, validate, and test our light curve classifier, we require a large set of light curve data for which classification has already been performed. In this paper, we utilize the light curve database and classifications provided by ASAS.

ASAS continuously monitors stars in the available sky for photometric variability in the \textit{V} and \textit{I} band. We choose the catalog of \textit{V}-band variables with 46744 light curves published 2002--2005 --- hereafter referred as ASAS 1.0. The light curves vary greatly in coverage, from less than 50 to more than 1000 measurements per star, but typically $\sim 220$ measurements over $\sim 3$ yr. For each periodic variable star the catalog provides the classification as determined by ASAS, the period, reference time for phase 0 and some ancillary information. The ASAS team published an updated catalog in August 2008 --- hereafter referred as ASAS 1.1 --- but we have opted to utilize the ASAS 1.0 database as its integrity has so far been more fully vetted.

Importantly, the ASAS database provides a classification for each light curve in which each object is identified as belonging to one of 13 classes of periodic variables.\footnote{Some ASAS classes (PULS, ACV, BCEP, CW-FU, CW-FO) were not used throughout the survey; we have opted to ignore these classes and have removed the 307 stars associated with these classes.}. These 13 light curve classes are described in Table \ref{classdistribution}. We adopt these same 13 light curve classes here. As described in Section \ref{sneuralnet}, the basic output of our classifier is a quantitative probability that a given light curve belongs to each of these defined classes.

\begin{table}
    \begin{tabular}{|l|r|r|l|}
        \hline
        Class Name & `Single Class' & `Clean'  & Type of Variable \\
        \hline
        DC-FO &   181  &   136  & $\delta$ Cepheid, first overtone    \\
        DC-FU &   579  &   388  & $\delta$ Cepheid, fundamental mode  \\
        DSCT  &   542  &   321  & $\delta$ Scuti                      \\
        EC    &  2661  &  2116  & over-contact eclipsing binaries   \\
        ED    &  1159  &   976  & detached eclipsing binaries       \\
        ESD   &   849  &   710  & semi-detached eclipsing binaries   \\
        MIRA  &  1592  &   156  & Mira type variables               \\
        MISC  & 24300  &  3448  & Miscellaneous                     \\
        RRAB  &  1225  &   470  & RR Lyra, type ab                  \\
        RRC   &   328  &   192  & RR Lyra, type c                   \\
        \hline
        Sum   & 33416  &  8913  & \\
        \hline
    \end{tabular}
    \centering
    \parbox{16cm}{
        \caption[Science Classes and Distribution]{\label{classdistribution} Science Classes and Distribution in the Single Class and Clean Data Set }}
\end{table}

There are some known issues with the quantities reported in the ASAS database that our light curve classifier must deal with. For example, a common issue with periods identified in an automated fashion is that in some cases the reported period is a harmonic of the true period, most commonly double of the true period. We accept this to be true for some of the reported ASAS periods, as well as any period finder we may use with our classifier in future. Thus the classifier needs to be trained to correctly identify an EB although it might be initially be associated with a harmonic of the true period, because we want the neural network to be able to adapt to data defects as long as the defects are part of the training.

The ASAS database is also known to contain some mis-classifications for various reasons. In particular, 11596 light curves or 25.7\% have more than one class reported in ASAS. As there is no probability provided in the ASAS database for distinguishing among the multiple classifications, they cannot be used for training purposes and are filtered out. The remaining 33416 light curves have only a single class assigned in the ASAS database and form what we call the 'single class' dataset.

Some light curves are too sparse, others too noisy or phased wrongly. Most of these types of ill-conditioned light curves can be avoided via a $\chi^2$ criterion. In particular, if we select only light curves for which the reduced $\chi^2$ of the polynomial light curve representation described in Section \ref{sneuralnet} is $\le 1.0$, we obtain a second data subset of 8913 light curves which we label 'clean'. The ASAS classes and the distribution of stars among the 'single' and 'clean' data subsets are given in Table \ref{classdistribution}. Note that the principal effect of filtering out ASAS light curves by the $\chi^2$ criterion is to eliminate the majority of the $\sim$24,000 objects that ASAS had classified as `miscellaneous'. We use both the 'single class' and the 'clean' data subsets for training and testing of our neural network classifier, as described in Section \ref{sneuralnet}.

\section{\label{sneuralnet} Light Curve Classifier: Neural Network Configuration, Light Curve Parameterization, and Training}

In this section we describe our neural network based light curve classifier. First we describe the adopted configuration of the neural network. Next we present a simple polynomial-based method for characterizing each light curve with a minimal set of shape parameters for input into the neural network. Finally, we discuss the training of the neural network using ASAS light curves characterized with this polynomial representation.

\subsection{\label{neuralnetconfig} Neural Network Configuration}

\begin{figure}[htb]
\epsfig{file=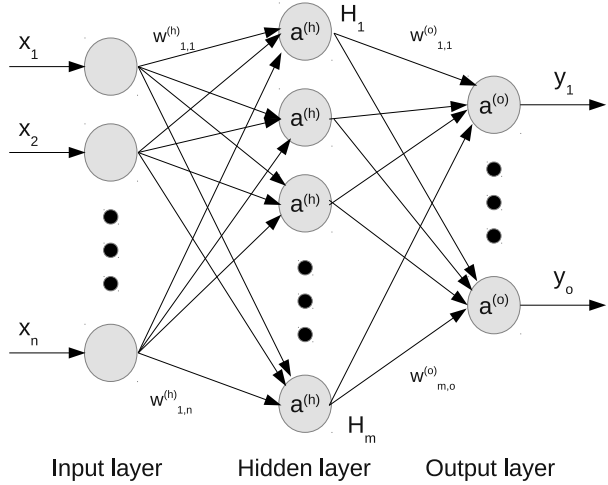, scale=0.50}
\centering
\parbox{16cm}{
\caption[Neural Network Architecture] {\label{neuralnet} Our neural network configuration follows a typical three-layer feed-forward neural network architecture with $n$, $m$ and $o$ neurons in the input, hidden and output layers, respectively. In our implementation, the input neurons correspond to the 19 shape parameters that we measure for each light curve (see Section \ref{polyfit}). Denoting layers by superscripts, $a^{(h)}, a^{(o)}$ are the activation functions of the hidden and output layer. $w^{(h)}, w^{(o)}$ are the weights connecting the layers. See Appendix \ref{appneuralnet} for the formal description of the hidden neurons, their weights, and the associated activation functions. In the output layer, $y_i$ are the output probabilities corresponding to each of the possible 13 light curve classifications (see Table \ref{classdistribution}). Not shown in the figure is the bias, an additional neuron for numerical stability with a constant value of $-1.0$, that is connected to each neuron in the input and hidden layer. }
}
\end{figure}

Artificial neural networks are simplified models of their biological counterparts and were originally designed to replicate the behavior of the human brain. The original idea dates back to \citet{annorig} and is represented in Figure \ref{neuralnet}. A set of input values (observables) are processed by "neurons" (one for each input). The neurons seeing the input data are connected to additional layers of so-called hidden neurons in various combinations. The last layer of neurons produces an output or result. In our case the sensory inputs are a set of $n = 19$ light curve shape observables defined in Section \ref{polyfit} (designated as $x_i$ in Figure \ref{neuralnet}), the output is a set of probabilities associated with each of the 13 possible light curve classes in Table \ref{classdistribution} (designated $y_i$ in Figure \ref{neuralnet}).

We defer the detailed formalism of our neural network configuration to Appendix \ref{appneuralnet}. Briefly, and in conceptual terms, whereas biological neurons are connected by synapses, the artificial neurons are connected by weight factors. Each neuron of the first layer multiplies every input, which in our case is each of the 19 light curve shape parameters described in Section \ref{polyfit}, by an individual weight factor. These weighted inputs are then passed to an 'activation function' which determines if the neuron in the hidden layer gets 'activated' or not. Through the activation of specific hidden layer neurons by the weighted inputs, specific output neurons are in turn activated by various amounts, resulting in our case in the output probabilities associated with each of the 13 possible light curve classifications. Therefore, the hidden layer neurons, and their associated weights and activation functions, are the process by which the neural network `transforms' the input signal (the 19 light curve shape parameters described in Section \ref{polyfit}) into the output signal (the light curve classification probabilities). In general, there may be multiple layers of hidden neurons, however for simplicity we have opted on a neural network configuration with a single hidden layer, as depicted in Figure \ref{neuralnet}.

The number of hidden layer neurons, and the values of the weights connecting them to the input and output neurons, are determined iteratively through a process of `training' that exposes the neural network to a large number of example cases for which the inputs and outputs are known. This is intended to simulate the process by which biological neural networks learn. A model for biological learning by biochemically modifying the electrical potential of the synapses was introduced by \citet{hebb}. In the artificial case this is modeled by adjusting the weights. The concept of modifying the weights by back-propagating error --- defined as difference between the output of a neural net and the correct result --- was introduced in \citet{backprop}. Our implementation of training the neural network for our light curve classifier  follows these precepts, as we now described below and in Appendix \ref{appneuralnet}.

\subsection{Polynomial Representation of Light Curve Shape as Input for Light Curve Classification \label{polyfit}}

To determine the probability that a given light curve belongs to one of the 13 light curve classes in Table \ref{classdistribution}, our neural network based classifier requires a set of input parameters that can be directly and uniformly measured for each light curve. Ideally these parameters should be small in number so as to enable rapid performance of the neural network, yet be sufficient in number to accurately characterize the most important distinguishing properties of the light curves. In contrast to previous approaches which have typically used Fourier series with as many as 30 parameters to describe the light curves (\citet{sarro09}; \citet{debosscher09}) we have opted for a geometric shape parametrization approach that is simpler computationally and requires only 19 parameters, as we describe in this section.

\begin{figure}[thb]
\epsfig{file=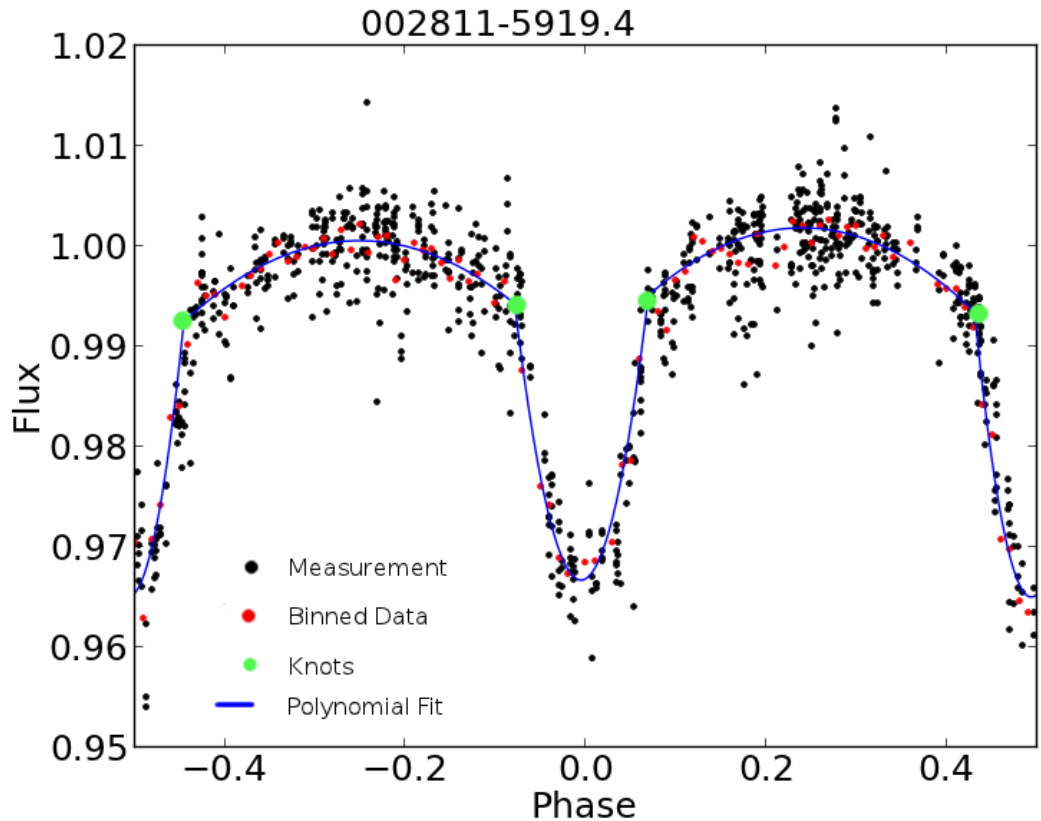, scale=0.75}
\centering
\parbox{16cm}{
\caption[Sample light curve] {\label{samplelc} Sample light curve of an eclipsing binary. Black: original measurements, red: binned data, blue curve: polynomial fit, green circles: position of the knots.}
}
\end{figure}

We represent each light curve with a simple multi-component polynomial fit, using the {\tt polyfit} algorithm of \citet{prsa08}, which can be computed quickly and easily. The original {\tt polyfit} algorithm as published in \citet{prsa08} fits a chain of 4 polynomials of a variable degree to a phase-folded light curve. An example is shown in Figure \ref{samplelc}. The chain of polynomials (blue) is continuous, but not differentiable at the knots (green) where the polynomials come together. In our implementation, we use four polynomials of second order (12 parameters) and we also leave the phase positions of the knots joining the polynomials (4 parameters) as free parameters.

\begin{figure}[thb]
\epsfig{file=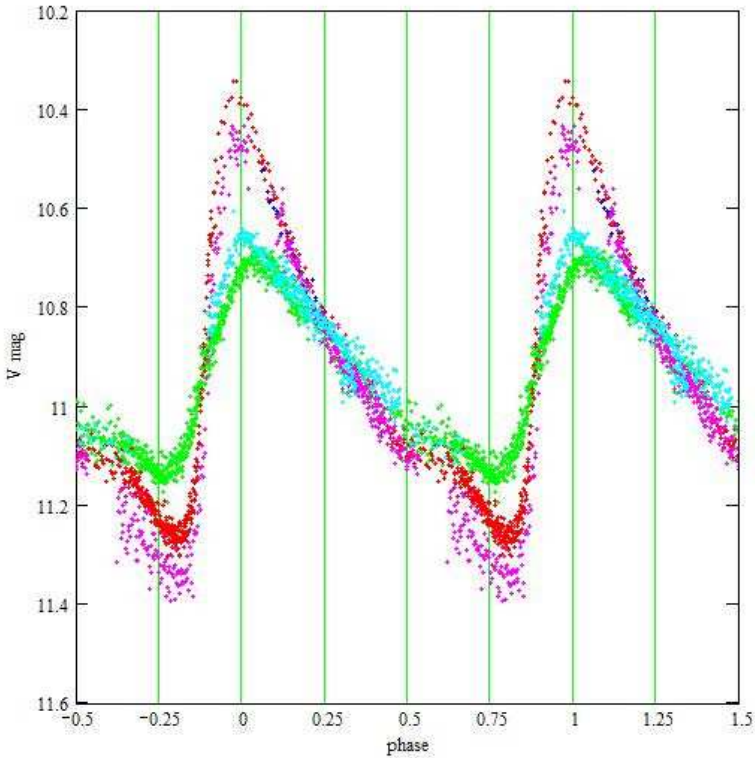, scale=0.75}
\centering
\parbox{16cm}{
\caption[Blazhko effect] {\label{blazhko} Light curve of RV Cet, a RR Lyra star with long-periodic magnitude changes (Blazhko effect, different Blazhko periods are color coded. Observation and figure by M. Bonnardeau, MBCAA Observatory, France.}
}
\end{figure}

The original polyfit algorithm searches for the first instance where the data points cross the mean value. This is the first knot and points are collected until they cross the mean again where the second knot is assumed. With noisy data this often leads to a broken chain. Binning the light curves is one way to deal with this problem and in our case solved the problem for light curves with more than 200 measurements. For less densely populated light curves we modified the algorithm to allow one isolated point to be on the other side of the mean value as long as the next point is on the same side again. This turned out to be a successful strategy. Binning does reduce high frequency noise in the original data, but destroys part of the signal, because for example RR Lyrae stars with Blazhko effect show slightly different light curves per Blazhko period (see Figure \ref{blazhko}). In order to keep at least part of this information, we compute the reduced $\chi^2$ value of the polyfit and use it as an additional input parameter for the neural network.

\begin{figure}[htb]
\epsfig{file=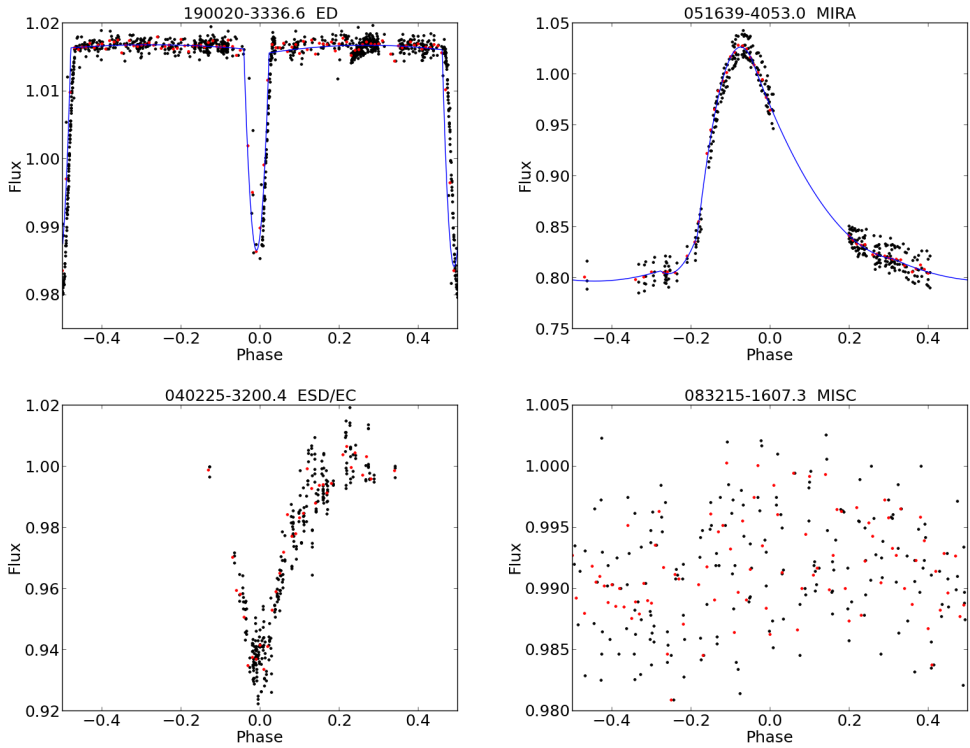, scale=1.00}
\centering
\parbox{16cm}{
\caption[Sample light curves] {\label{lcsamples} Sample light curves: black dots are ASAS data, red dots are binned data, the blue line is the fitted polynomial. Top row: successful fits. Bottom row: fitting failed due to insufficient phase coverage (left) and too much noise (right).}
}
\end{figure}

A small number (1736) out of the 46749 ASAS light curves or 3.7\% could not be successfully fitted by our modified polyfit algorithm. Some examples are shown in Figure \ref{lcsamples}. Most of the failed fits were due to insufficient coverage in phase and/or noisy light curves. About 15\% of the failed fits were Mira stars which are better matched by fitting 2 instead of 4 polynomials. We therefore modified polyfit to try both, a chain of 2 and 4 polynomials and then selected the one with the smaller $\chi^2$ value. In our final implementation, where two polynomials are a better representation as measured by $\chi^2$, we set the coefficients and knot positions for the third and fourth polynomials to 0.

Thus the input parameters for our neural network are the four knot coordinates for the different sections of our polynomial chain and the three coefficients for each of the second order polynomials. These 16 parameters describe the geometry of the light curve. As physical parameters we also include the logarithm of the period in days, the $\chi^2$ value of the polynomial fit, and the maximal amplitude of the light curve (in normalized flux units). The period and amplitudes are an obvious choice, for example while Mira variables have periods over 80 days and amplitudes of $2.5$ - $11$ mag (in \textit{V}), RR Lyrae stars have much shorter periods and smaller amplitudes (less than 2 days, less than 2 mag). The $\chi^2$ value is a less obvious choice, but as mentioned above we add it in order to give the neural network information on the extent to which the light curve deviates from our simple polynomial representation and thus gives information about higher-order light curve structure, such as in RR Lyrae with Blazhko effect (see Figure \ref{blazhko}), without necessitating computationally expensive Fourier series.

\subsection{Training the Neural Network}

The goal of the neural network training step is to expose the neural network to a large set of light curves that are representative in terms of cadence, noise, and other characteristics of the light curves to be classified, and whose classifications are already known, so that the hidden layer neurons and the weights connecting them to the input and output layers (Section \ref{neuralnetconfig} and Appendix \ref{appneuralnet}) can be determined and optimized. Once trained, the neural network can be fed the shape parameters for any new light curves (i.e., Section \ref{polyfit}), and the output classification performed extremely rapidly.

In order to get a numerically stable training and network, the input parameters have to be normalized. Arranging input values as columns and different light curves as rows, this usually means to compute the mean value and standard deviation of a column, say period, over the whole data set and then to normalize each value by subtracting the mean and dividing the remainder by the standard deviation. Any unknown light curve to be classified has to be normalized the same way as the training set before it is fed to the network. We apply this normalization to the physical parameters, log(period), and mean magnitude.

We chose a different strategy to the geometry parameters (the polynomial fit parameters). The knot coordinates do not need further treatment, because they are already between phases $-$0.5 and $+$0.5, a numerically stable interval. The polynomial coefficients are different, because they vary widely and over a range that would make any training numerically unstable. The usual normalization strategy of subtracting the mean does not preserve the sign of the coefficients and thus does change the form of the fitted polynomial. We implement a simple division by $f \sigma$ where $f$ is a free scaling constant and sigma is the standard deviation, thus preserving the sign of the polynomial coefficients.

As described in Section \ref{neuralnetconfig}, we choose to configure the neural network with just one hidden layer, but varied the number of neurons from 1 to 30. We expect networks to perform best, if the number of hidden neurons is about the same as the number of input parameters. If networks with much fewer hidden neurons perform better or as well, the network works as 'data compressor', much like loss-less classical compressing algorithms do on text files or images. This would be an indication of having more input variables than needed. We do not expect networks with more hidden neurons than input parameters to perform better. Instead the performance should flatten out, because the network does not learn anything new; adding more neurons than needed will increase the error and thus decrease the efficiency. For each number of hidden neurons we trained 10 networks --- training more networks did not show any significant difference.

For each of the training runs we applied a cross-fold validation technique. We randomly split the whole dataset into three subsets: training ($50 \%$), validation and testing ($25 \%$ each). The training set was used to train the network by letting it process this set 10 - 100 times. In order to avoid the network learning patterns related only to the sequence of the training set, we randomized the order of the training light curves after each of these runs. We then took the validation set --- the network did not see any light curve in the validation set during the training --- and let the network process it and computed the error in classifying the validation set. We repeated until the validation error starts to rise, which means that the network is trained. Finally we let this trained and validated network classify the testing subset and measured the retrieval rate and the false-positive rate.

For any given number of neurons in the hidden layer we repeat this process 10 times. Depending on the number of neurons in the hidden layer, training times vary between a few minutes and about 4 hr on a standard workstation for the same momentum and learning rate. Due to the fact that the initial weight factors chosen randomly, the training time can vary up to a factor of 2.5 even if the number of neurons are the same. Once trained, the classification of even the larger 'single class' data set is done in a few seconds on a standard workstation; in fact writing the results takes longer than the actual classification. The trained network with the best retrieval rate is chosen as the final classifier and its results are reported in the next section.

\section{\label{results} Results}

We have created and trained a neural network based light curve classifier utilizing simple geometric light curve shape parameters as described in Section \ref{sneuralnet}. Thus a basic result of this work is the classifier itself, with a ready-to-use trained neural network and ancillary programs for determining light curve shape parameters that can be applied to any light curves of interest. We provide the trained neural network, the ancillary programs, and a guide to their use in Appendix \ref{tools}.

In the following we discuss the performance of the classifier against the ASAS database. For each of the classes in Table \ref{classdistribution} our classifier produces a probability as output. The class with the highest probability is taken as result of the classification. For any given class the 'retrieval rate' --- our primary metric --- is defined by the ratio of the number of correct classifications to the true total number of light curves in this class.

\subsection{\label{sectsingleclass} Classification of 'Single Class' Light Curves}

\begin{figure}[htb]
\epsfig{file=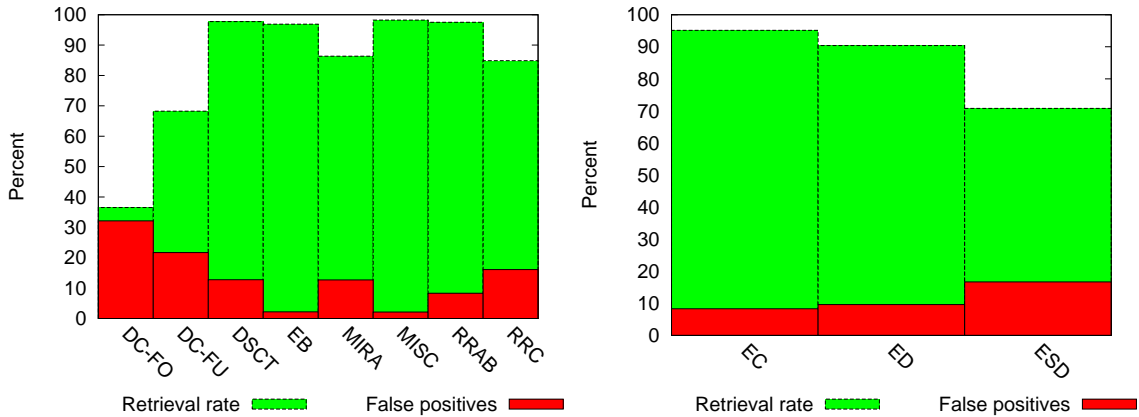, scale=0.75}
\centering
\parbox{16cm}{
\caption[Classification Results Single Class Variables] {\label{singleclassresult} Left: retrieval and false-positive rates for variables having a single class specifier in ASAS. Right: same as left, but for EB subclasses.}
}
\end{figure}

We trained networks with $1$ to $30$ hidden neurons, the best performing network has 19 hidden neurons. Aggregating detached, semi-detached and over-contact EBs into one class, Figure \ref{singleclassresult} gives an overview of the retrieval and false-positive rates for EBs and the other classes given in the left panel of Table \ref{classdistribution}. The right panel shows the retrieval rate and false positives for EB subclasses. The detailed confusion matrix showing the distribution of correctly and incorrectly classified light curves is in Table \ref{singleclasscm} in Appendix \ref{appendix}.

While we achieve good results for EBs overall ($96.9 \%$ retrieval rate, $2.2 \%$ false positives), the performance is less robust for the EB subclasses (ED, ESD and EC in Table \ref{classdistribution}). Therefore we next investigate the subclasses in greater detail. 

\begin{figure}[htb]
\epsfig{file=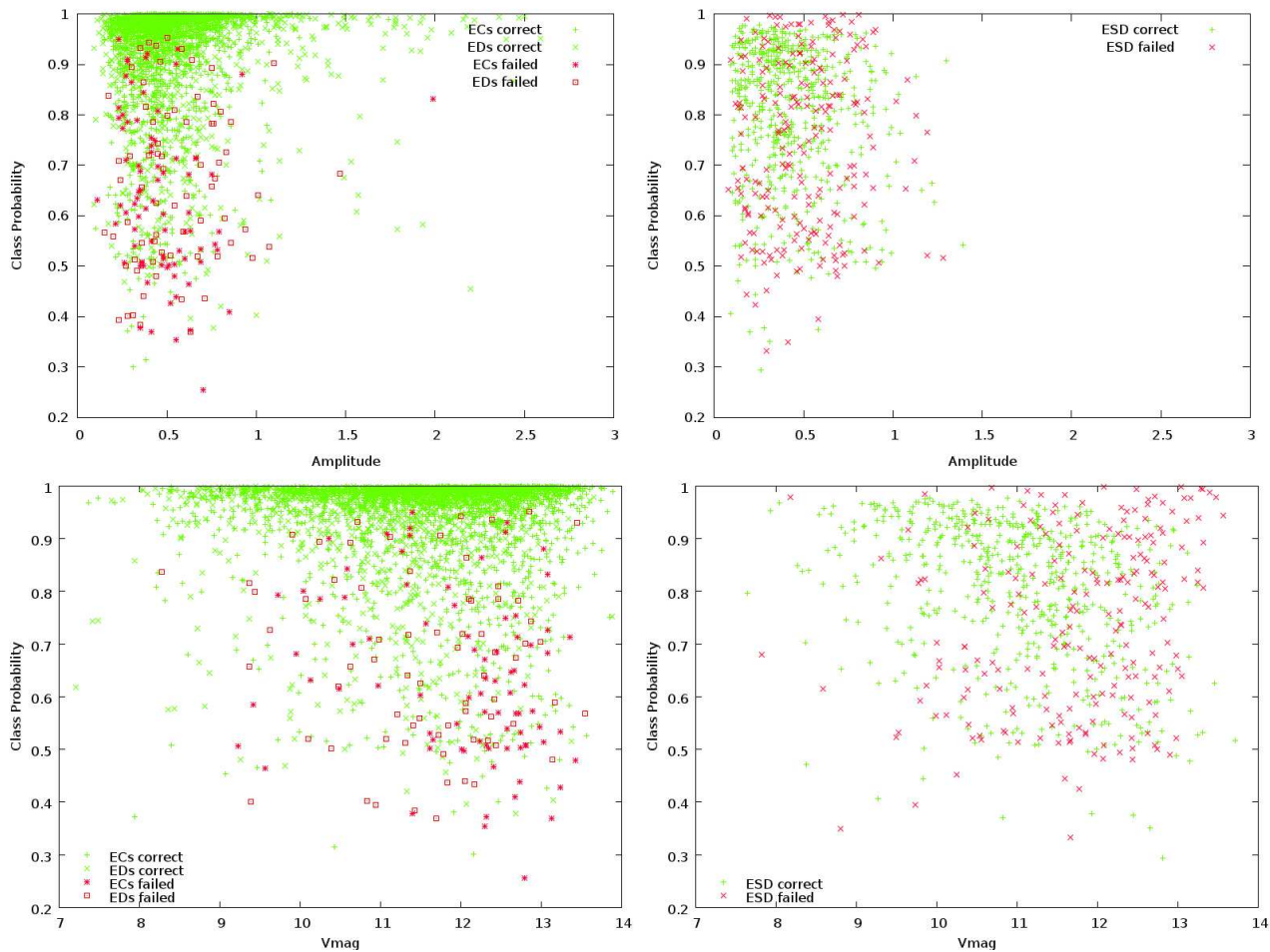, scale=1.25}
\centering
\parbox{16cm}{
\caption[Distribution of EBs] {\label{ebsdistro} Probability distribution of correctly and mis-classified EBs over amplitude and \textit{V} magnitude. Left column: detached and over-contact systems, right column semi-detached systems.}
}
\end{figure}

The ASAS classification uses ratios of amplitudes of a Fourier-series decomposition of the light curve for sub-classifying EBs, and this classification becomes very uncertain if the amplitude ratios are small (\citet{asasebs}). In Figure \ref{ebsdistro} we plot the membership probability as computed by our network for correctly and mis-classified EBs versus the amplitude in \textit{V} and the magnitude itself. The upper panel shows that mis-classifications for all EBs are fairly evenly distributed for amplitudes less than $1.0 \mbox{mag}$. For detached and over contact systems (upper left) all classifications with a probability $p > 0.95$ are correct, no matter how shallow the amplitude. While the density of misclassified systems (red symbols) is fairly even in the range of $[0 - 1.0; 0.35 - 0.95]$, the mis-classification ratio (red to green symbols) rises sharply for $p < 0.85$. Therefore the detached and over-contact EBs appear well conditioned as long as the classifier yields $p \ge 0.85$.

In contrast for semi-detached systems (upper right) there is no probability above which all classifications are correct, in fact the systems with the highest probabilities are all mis-classified ($90 \%$ of them as over-contact systems). \citet{asasebs} finds that the sub-classification for EBs improves for large amplitudes, but unfortunately we find that the classification of semi-detached EBs does not improve with lager amplitudes.

In addition, plotting the probability against the \textit{V}-magnitude (lower row of Figure \ref{ebsdistro}) we find that the semi-detached systems (lower right panel) become even less well classified at faint \textit{V} magnitudes, whereas detached and over-contact systems do not suffer from this effect. Thus it appears that our classifier is quite robust for detached and over-contact EBs, but struggles with semi-detached systems.

We find that the semi-detached systems are twice as likely to be confused for over-contact EBs than for detached EBs (see Table \ref{singleclasscm}). Fortunately for the EB Factory project the mis-classification of semi-detached as over-contact EBs is not relevant, because we plan to analyze both sub-classes together in the following step of the analysis pipeline. The mis-classification of semi-detached systems as detached is potentially problematic, especially because detached EBs are generally the most useful for detailed follow-up study.

We tried to resolve the problem of semi-detached systems using the 'Local Linear Embedding' method reported in \citet{lle} and successfully tested with \textit{Kepler} EBs. Using LLE we re-classified all EBs and repeated training, validation and testing. Unfortunately we found no significant improvement. We are continuing to explore alternative solutions, but sub-classification of semi-detached EBs continues to be a challenge.

\begin{figure}[htb]
\epsfig{file=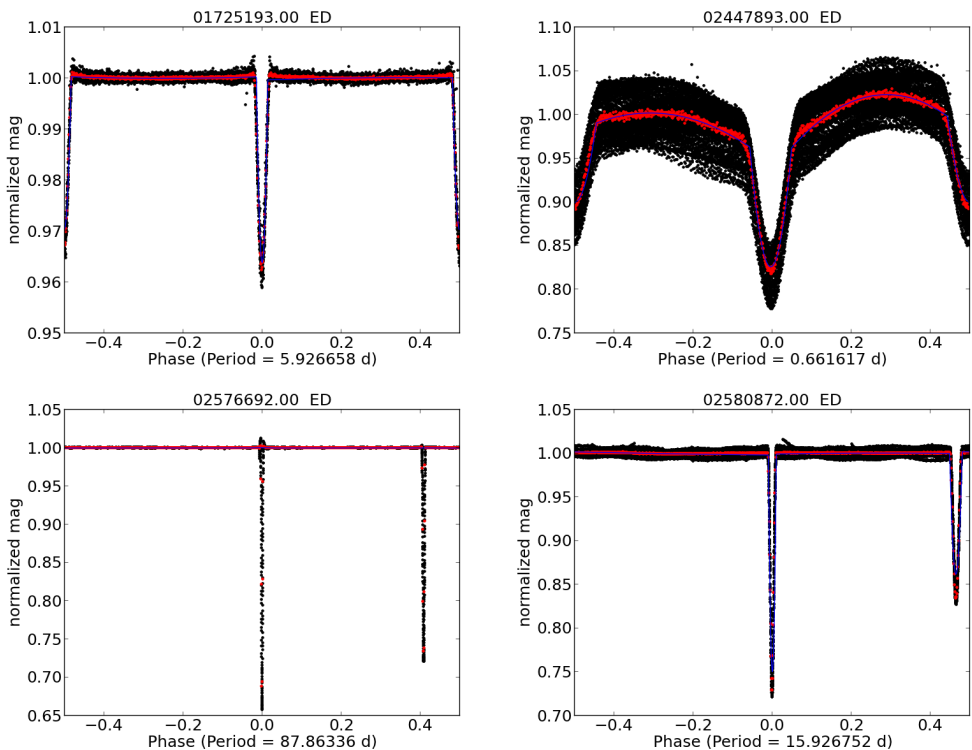, scale=1.0}
\centering
\parbox{16cm}{
\caption[Examples for \textit{Kepler} light curves] {\label{keplcs} \textit{Kepler} light curves: Top left: successfully retrieved ED; top right: ED re-classified as ESD; bottom left: eclipses missed by polyfit; bottom right: ED classified as MISC due to narrow eclipse.}
}
\end{figure}

As a final test of our classifier we took the \textit{Kepler} eclipsing binaries published in \citet{keplerebs} and let the network classify them. Counting retrieval as EBs, the rate for ECs is 98\%, for ESDs 90\% but for detached binaries it fell to a surprisingly low 59\%; see Table \ref{keplercm} in Appendix \ref{appendix}. A closer inspection revealed that almost half of the detached binaries were classified as MISC for two reasons: (1) wide binaries with an extremely narrow, sharp drop in the main eclipse are under-represented in the ASAS data set due to the much higher photometric precision of \textit{Kepler}, and (2) many narrow, deep eclipses were missed by our polynomial fitter. Figure \ref{keplcs} shows a few example light curves. For future classifications we will have to improve the polynomial fitter for narrow eclipses and to include real or simulated wide binaries into the training, validation and test sets. However, this preliminary test shows, that our classifier should perform extremely well with high precision light curves such as \textit{Kepler}'s.

\subsection{\label{sectcleanlc} Classification of 'Clean' Light Curves}

In order to assess the impact of light curve quality on the performance of our classifier, we trained, validated and tested our classifier on this smaller but higher quality data set in the same manner as the 'single class' data set. The results are summarized in Figure \ref{cleanresult}, the confusion matrix is in Table \ref{cleancm} in Appendix \ref{appendix}.

\begin{figure}[htb]
\epsfig{file=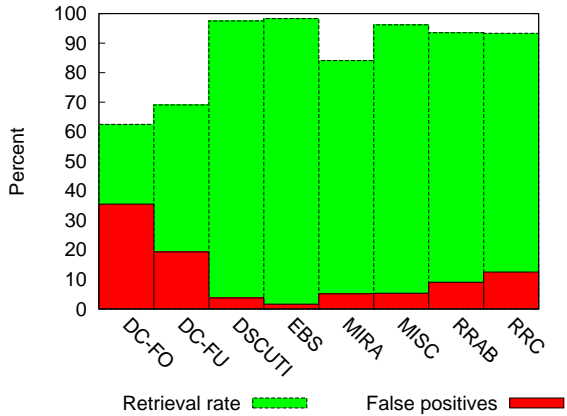, scale=0.75}
\centering
\parbox{16cm}{
\caption[Classification Results Variables with Clean Light Curves] {\label{cleanresult} Retrieval and false-positive rates for variables having a single class specifier in ASAS and $\chi^2 \le 1.0$. Result for EBs: 98.3\% retrieval rate, 1.6\% false positives.}
}
\end{figure}

Aggregating EBs into one class, we get $98.3 \%$ retrieval rate and $1.6 \%$ false positives which is slightly better than for the 'single class' data set.  Except for $\delta$-Cepheids with overtones (class DC-FO) the difference between the 'single class' and clean data set is in the order of $1 \sigma$ for different runs involving randomized training, validation and testing set. However, the sub-class confusion for EBs shows the same characteristics as described above for the 'single class' data set.

\begin{figure}[htb]
\epsfig{file=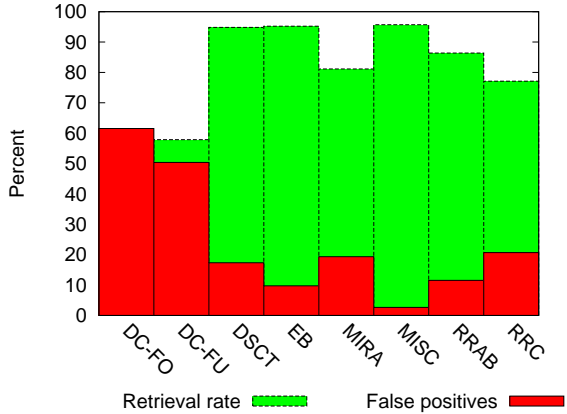, scale=0.75}
\centering
\parbox{16cm}{
\caption[Classification Results single class dataset trained with clean light curves] {\label{singlewithclean} Retrieval and false-positive rates for variables having a single class specifier in ASAS by a classifier trained with clean light curves. Result for EBs: $95.2 \%$ retrieval rate, $9.8 \%$ false positives.}
}
\end{figure}

We ran the network trained with the much smaller, clean data set over the whole single class set. The overview is in Figure \ref{singlewithclean}, the confusion matrix in Table \ref{singlewithcleancm}. The retrieval rate for EBs declined from $98.3 \%$ to $95.2 \%$, taking EBs as one class. The rate of false positives went up from $1.6 \%$ to $9.8 \%$. Interestingly it appears that training with clean light curves leads to an increased mis-classification of the full data set because the network was not exposed to enough examples of noisier light curves. The main contributors are stars with an ASAS class of RRAB and MISC. RR-Lyrae should be better separated from eclipsing binaries with color information, which we will add in future improvements of the classifier. 

Finally we applied this network to the \textit{Kepler} EBs and produced similar results as reported above. Evidently the classifier trained only on 'clean' light curves does not suffer from lack of exposure to noisy light curves, because the \textit{Kepler} light curves have very low noise.

\section{\label{discussion} Discussion}

Our results presented above have focused on the performance of our classifier for EBs and their sub-classes, as this is the primary intent of the EB Factory project. However, it is very useful to consider the performance of the classifier more generally across all variability classes.

Taking a closer look at the confusion matrices (see Appendix \ref{appendix}), we do quite well for eclipsing binaries, RR Lyra and Delta Scuti stars (see also Figure \ref{singleclassresult}). On the other hand the retrieval rates for multi-mode Cepheids (DC-FO) are very low. In part this is due to the fact that phase folded light curves of variables with multiple modes are not well represented by a polynomial fit. We had hoped to overcome this by including the $\chi^2$ residual of the polynomial fit. This improved the retrieval rate for RR Lyrae showing Blazkho variability, but it did not help as hoped for multi-mode variables. One possible way to deal with this problem is to add the number of periods found or at least the secondary period as one additional input parameter. 

Another source of uncertainty lies in the ASAS data set itself. For EBs \citet{richards12} found wrong periods in $29 \%$ of all cases. Visual inspection showed aliased periods for other classes as well. A second problem revealed by inspecting the light curve is incorrect phasing. EBs are quite often not phased on the primary eclipse and Pulsators are not phased on their maximum brightness. If the phasing is not correct, the primary eclipse minimum lies at a phase other than 0. This phase shift of the minimum has a serious impact on the retrieval rate for our classifier. In our current implementation we simply used the phase for the minimum given by ASAS, however we found that zero point is not always correct. Although the retrieval rate for EBs is high, we expect to improve it further by re-phasing the ASAS light curves consistently.

Any machine learning classification is only as good as its training-set and using the ASAS dataset has its caveats. It does have the advantage to provide a substantial number of light curves for nine scientific classes of periodic variables. The tenth class however - MISC - is by far the strongest in numbers and seems to have served as a kind of "light curve trash can". Our network seems to have learned this logic well enough and consequently has classified EBs wider than ever seen before as MISC. \citet{richards12} put a lot of effort into beating down the high number of miscellaneous variables and re-classified the ASAS 1.1 data set. However, no automatic classification is free of errors and while \citet{richards12} gives a number of about $24 \%$ for the mis-classifications in the original data set, they give a careful 'less than $20 \%$' for the re-classified catalog. 

Subclass-confusion for eclipsing binaries is a problem for all classifiers. \citet{sarro09} does not report EB subclasses when applying the classifiers to the OGLE database, neither does \citet{blomme11}. \citet{hippcls} reports results similar to ours. \citet{richards12} gives numbers for corresponding classifications between ASAS and their random forest classifier with and without outlier rejection (Figure 13 therein), the results without outlier rejection are comparable to ours. Retrieval rates vary between $50 \%$ (\citet{sarro09}) and $98.3 \%$ (\citet{blomme11}), matching our best result.

\section{\label{summary} Summary}

Using the ASAS database we trained a neural network to classify periodic variables using a simple parameterization of the light curve geometry by a polynomial fit. We achieve retrieval rates up to $98.3 \%$ for eclipsing binaries and similarly high retrieval rates for most other classes. Our network uses many fewer input parameters than previous approaches. We keep the $\chi^2$ of the fit as an input parameter in order to help distinguish between single and multi-mode variables with an improvement for RR Lyrae variables with Blazhko effect, but less effective for multi-mode variables such as first overtone Cepheids. We expect to improve the network performance for multi-mode variables by including at least one additional period as an additional input parameter. 

Running the trained network on the known \textit{Kepler} EBs we find that ASAS does not include enough wide EBs with narrow and steep eclipses and thus our ASAS trained classifier fails for these. For the narrowest eclipses we will have to improve the polynomial fitting, because they are missed by the current algorithm. We might further improve the results by switching to a different set of geometry parameters that do not require normalization, because our current normalization scheme results in a loss of some light curve shape information.

For eclipsing binaries the sub-classes (detached, semi-contact, over contact) are not adequately separated. We plan to try a refined Local Linear Embedding scheme as in \citet{lle} to get a better separation between detached and contact systems. However, in truth the difficulty in separating the EB subclasses is driven almost entirely by the semi-detached systems. This is likely to continue to pose a challenge. At the same time, if the primary goal is simply to identify EBs as general class, and to distinguish them from other variable classes, then our current classifier already performs extremely well. 

An important limitation of our classifier, or any other light curve classifier, is that it relies strongly on the accuracy of the period and the zero phase of the light curve, because it is trained to expect key features of the light curve to occur at certain phases (such as the primary eclipse at phase 0).

An important lesson of the investigation presented here is that it is crucial to train the classifier with light curve exemplars that include similar noise characteristics as the light curves it is intended to classify. Our classifier trained on only the very best ASAS light curves performs less well on the full data set than when trained with the full range of light curves. Because our implementation uses a much simplified representation of the light curves and only a couple of additional parameters, such as the $\chi^2$ of the polynomial representation, it involves $\sim 30 \%$ fewer parameters than previous classifiers. Moreover, these geometric light curve shape parameters are very easily and rapidly computed, in contrast to Fourier based methods, resulting in a substantial speed-up of the light curve characterization and the neural network training and validation steps.

We have trained and vetted our neural network classifier on the ASAS data set, and we provide the trained neural network along with ancillary programs for light curve characterization and a guide for use in Appendix \ref{tools}. These tools should enable the community to perform rapid, bulk classification of large numbers of light curves from current and upcoming all-sky time-domain surveys.

\acknowledgments
This work has been funded by the NASA ADAP grant NNX12AE22G. We thank Dr.\ Joshua Pepper and Dr.\ Nathan De Lee for valuable discussions and help.

\begin{appendices}

\section{Neural network geometry \label{appneuralnet}}

In the following equations we indicate the hidden and output layer by the superscripts $^{(h)}$ and $^{(o)}$. For a given input vector $x_i$ and an activation function $f^{(h)}$ for the hidden layer we successively compute the activity level $a^{(h)}_j$ for each of the hidden neurons $H_1$ to $H_m$: 
\begin{equation}
a^{(h)}_j = f^{(h)} \left( h^{(h)}_j \right) \quad \mbox{where} \quad h^{(h)}_j = \sum_i w^{(h)}_{i,j} x_i.
\end{equation}
We repeat the same operation for getting the outputs $y_1$ to $y_o$, taking the activation levels we just computed as inputs and substituting the hidden layer activation function with the activation function of the output layer $a^{(o)}$:
\begin{equation}
y_k = a^{(o)} \left( \sum_j w^{(o)}_{j,k} a^{(h)}_j \right).
\end{equation}
As activation functions we chose the logistic function for the hidden layer
\begin{equation}
a^{(h)}(x) = 1 / (1 + \exp(-\beta x)),
\end{equation}
where $\beta > 0$ is a scaling parameter that is kept constant for a network, in our case $\beta = 1.0$. For the output layer we use the softmax function:
\begin{equation}\label{eqoutvals}
p_k = y_k = a^{(o)}(h^{(o)}) = \frac{\exp(h^{(o)}_k)}{\sum_{j = 1}^{o} \exp(h^{(o)}_j)} \quad \mbox{, where} \quad h^{(o)}_k = \sum_j w^{(o)}_{j,k} a^{(h)}_j
\end{equation}
$p_i$ is the normalized probability of $i$-th output neuron: $\sum p_i = 1$. 

Finally we compute the error of the outputs using the known values of the training set $t_i$
\begin{equation}\label{eqerror}
E = \frac{1}{2} \sum_k \left( t_j - y_k \right)^2.
\end{equation}

The network 'learns' by adjusting the initially randomly chosen weights in each layer in order to minimize the error. Seeing the error as a function of the weights we apply a downhill-gradient optimization and propagate adjustments to the weights backward through the net. Taking $n$ and $n + 1$ as subsequent time-steps the update rule for the weights is
\begin{equation}\label{equpdates}
\left( w^{(o)}_{j,k}\right)^{n+1} = \left(w^{(o)}_{j,k}\right)^{n} - \eta \frac{\partial E}{\partial w^{(o)}_{j,k}} + \mu \Delta \left(w^{(o)}_{j,k}\right)^{n} \qquad ; \quad
\left(w^{(h)}_{i,j}\right)^{n+1} = \left(w^{(h)}_{i,j}\right)^{n} - \eta \frac{\partial E}{\partial w^{(h)}_{i,j}} + \mu \Delta \left(w^{(h)}_{i,j}\right)^{n}.
\end{equation}
The first term on the right-hand side is the value of a weight in the current time-step, the second term the gradient of the error $E$ with respect to this weight multiplied with the 'learning rate' $\eta$. The learning rate is a numerical constant - usually with $0 < \eta \le 1$ - used to stabilize the learning against noise or over-compensation of errors in the input data. The third term connects changes made in the current learning step to the updated value for a weight factor, the controlling constant $\mu$ is called 'momentum'. Increasing momentum prevents to get stuck in a local minimum and accelerates the learning process.

Substituting the output values $y_k$ in Equation \eqref{eqerror} by Equation \eqref{eqoutvals} and applying the chain rule we get the gradient for the output layer weights
\begin{equation}
\frac{\partial E}{\partial w^{(o)}_{j,k}} = \frac{\partial E}{\partial y_k} \frac{\partial y_k}{\partial h^{(o)}_k} \frac{\partial h^{(o)}_k}{\partial w^{(o)}_{j,k}} = (t_k - y_k) \cdot y_k (1 - y_k) \cdot a^{(h)}_j
\end{equation}

The gradient for the hidden layer weights follows by applying the chain rule one step further:
\begin{equation}
\frac{\partial E}{\partial w^{(o)}_{j,k}} = \sum_k \left( \frac{\partial E}{\partial y_k} \frac{\partial y_k}{\partial h^{(o)}_k} \sum_l \left( \frac{\partial h^{(o)}_k}{\partial h^{(h)}_l} \frac{\partial h^{(h)}_l}{\partial w^{(h)}_{i,j}} \right) \right) =
a^{(h)}_j ( 1 - a^{(h)}_j) x_i \sum_k w^{(o)}_{j,k} (t_k - y_k) y_k (1 - y_k)
\end{equation}
Together with Equation \eqref{equpdates} the last two equations complete the learning by propagating the error backward. In summary the back-propagation process involves running a neural network over an input data set by feeding the inputs forward to the hidden and further to the output layer (Equation \eqref{eqoutvals}). These are compared to known outputs, and the differences define the errors. Then the error is computed according to Equation \eqref{eqerror} and propagated backwards to update the weights for the hidden and output layer (Equation \eqref{equpdates}). This pattern can be easily extended to any number of hidden layers.

\section{Software tools and database structure \label{tools}}

In course of this project we developed software for batch processing 50,000 - 160,000 light curves. For rapid prototyping and easy modification we use Python 2.7 as programming language and SQLite as a database. Although as interpreter language native Python is much slower than compiled C-programs, this does not play a big role in our case, because we are limited by disk input and output, not by CPU-time. There is one exception: training the network, for which we use a parallel algorithm and let it run on the ACCRE cluster at Vanderbilt. We use speed optimized modules (\texttt{numpy}) but expect that coding in Cython or C++ might further speed up the training process.

We choose SQLite as database. It is easy accessible from Python and C++, it allows very rapid development, does not require to set up a server and databases are easily copied between different machines. On the downside we experienced a drastically increased access time if more than one table is stored in a SQLite file and if the file exceeds about 1 GB in size. We decided to keep big tables in their own file; this allows to switch between different phasing algorithms by having one database file for each, on the other hand it means we have to enforce database consistency by our software. SQLite will be replaced by MySQL/MariaDB when running over the full \textit{Kepler} data set. The database access is encapsulated in classes, so the switch can be done without changing the higher level source code.

\begin{figure}[htb]
\epsfig{file=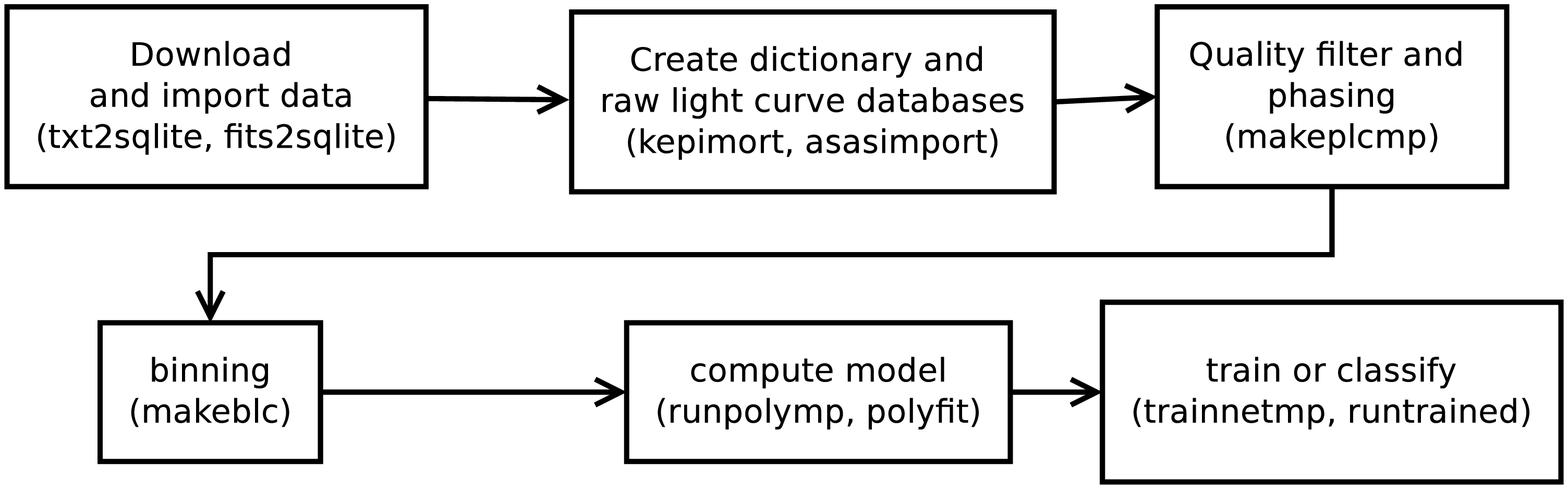, scale=0.33}
\centering
\parbox{16cm}{
\caption[General work flow] {\label{workflow} After download the information is converted into a database: stellar dictionary and raw light curves. The raw filtered by quality flags if available, then normalized, phased and binned. The binned light curves are then processed by polyfit which results in a model description which is then used to train the network or to let it classify unknown light curves.}
}
\end{figure}

The general work flow is shown in Figure \ref{workflow}, the data flow is shown in Figure \ref{dataflowdia}. This Appendix will briefly describe the software and database architecture.

\subsection{Database Structure}

\begin{figure}[htb]
\epsfig{file=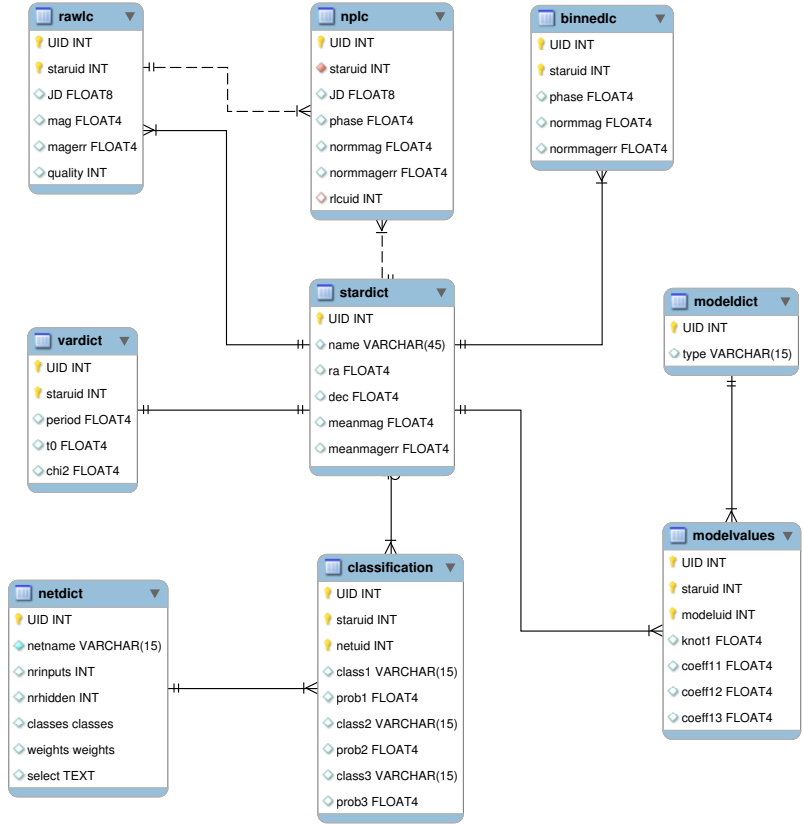, scale=0.75}
\centering
\parbox{16cm}{
\caption[Simplified database structure] {\label{eerdia} The star dictionary \texttt{stardict} at the center stores persistent values like coordinates, (mean or median) magnitudes and their errors. The variable dictionary stores variable stars with their characteristic values like periods, zero-point for the phased light curve and the quality of the fit (chi2). Time series are stored as raw light curves, which are filtered by quality, normalized and phased to the main period (nplc). The normalized and phased light curve is binned and serves as input for polyfit, resulting in model values. These values serve either for training - in case the star has been classified previously - or as input for a classification by a trained network.}
}
\end{figure}

\begin{figure}[htb]
\epsfig{file=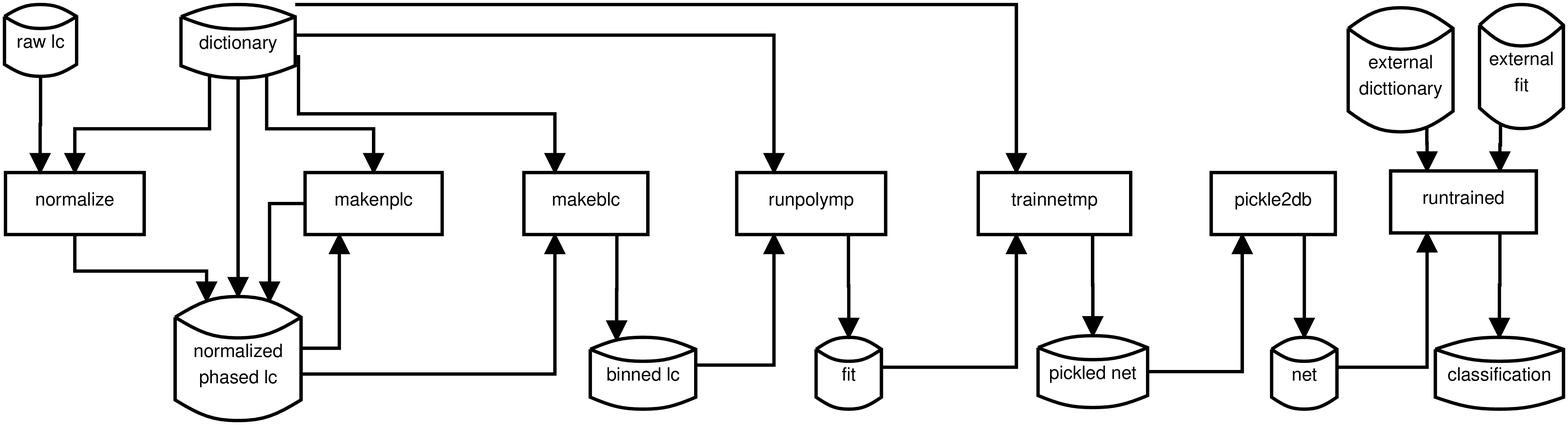, scale=0.33}
\centering
\parbox{16cm}{
\caption[Data flow] {\label{dataflowdia} Data flow diagram. The database for the dictionary and raw light curves are assumed to exist (different download formats make it difficult to write a general purpose import program). The raw light curves are selected by quality - if available - normalized and stored in their own database. \texttt{makeplc} phases the normalized light curves, \texttt{makeblc} bins them. The binned light curves are handed to \texttt{runpolymp} which generates a fit for each light curves. This fit is then either used to train a network (\texttt{trainnet}) or to have an already trained network to get a classification (\texttt{runtrained}).}
}. 
\end{figure}

Figure \ref{eerdia} shows a simplified model of the underlying database structure. Center piece is the stellar dictionary (stardict) which holds non variable information like coordinates, mean or median magnitudes and errors or keys to other databases like the Two Mircon All Sky Survey (2MASS). The variable stars dictionary (vardict) holds basic information for variable stars like periods, zero offsets for the main period and quality measures for the fitted model of the phase folded light curve (chi2). Currently the variable dictionary holds as well the previously known classification and---if available---the probability for this class. The tables for stars and variable dictionaries are stored in one database file, further referenced as the dictionary.

The downloaded, raw light curves are stored in "rawlc" - together with flags indicating the quality of the measurement. The table "nplc" is a subset of "rawlc", filtered by quality, with a phase added for periodic variables and with normalized magnitudes or fluxes. The normalized and phased light curve is finally binned (binnedlc). Most important: the stellar dictionary provides a unique ID for every star (UID), basically an unsigned number. This number is referenced as "staruid" in all other tables linked to a specific star. In database terms this is a "foreign key constraint", meaning that every entry in a light curve table must reference a stellar UID. This way the referential integrity of the database is granted.

We are currently able to use three different models for training a network or classifying a light curve: the coefficients of the polynomial fit, the knots and midpoints of fitting polynomials and a "full fit" consisting of an arbitrary number of equidistant phases and values for that phase. We separate between different fits by using the model dictionary. However, in this paper only one model - coefficients - is described and thus shown in Figure \ref{eerdia}. The tables "modeldict" and "modelvalues" together are the fit database.

"netdict" is the dictionary for trained networks, ready for classification. We keep a name for easy reference and the select statement we used to create the data set for training. In addition we store all weights and network internal information like learning rate and momentum.

Finally the classification table holds our own classification (pairs of class and probability to be exact). The referential integrity is again guaranteed by  "staruid", referencing the unique ID assigned to a star. Thus we can easily compare our to any other, previously known classification stored in the dictionary for variables.

While Figure \ref{eerdia} only shows a part of the database columns, the complete schema is currently stored in Python classes, so that the tables will create themselves if they do not yet exist.

\subsection{Walkthrough}

We use two helper programs (\texttt{txt2sqlite} and \texttt{fits2sqlite}) to convert the initial downloads into databases. However, quite often we need a customized program to split these into a dictionary and raw light curves. Once we have the dictionary and raw light curve database, the rest is straight forward. All subsequent programs are controlled by command line options, the most common ones are given in Table \ref{cmdopt}.

\begin{table}
	\begin{tabular}{|l|l|l|}
		\hline
		Option   & Default & Description \\
		\hline
		h        & None   & Help text with all available options \\
		rootdir  &   .    & Directory for database files \\
		dict     & None   & Stellar dictionary \\
		select   & select * from stars & Select statement \\
		selfile  & None   & File containing select statement \\
		d        & 1      & Debug flag \\
		dbconfig & Asas   & Name for the database specific configuration \\
		\hline
	\end{tabular}
    \centering
    \parbox{16cm}{
        \caption[Common command line options]{\label{cmdopt} Most Common Command-line Options}}
\end{table}

The raw light curves are processed by a series of batch programs. We take ASAS as sample database in this section. The steps can be executed and the results checked individually or they can be aggregated by a shell script. The first step is to do a quality selection and to normalize the magnitudes:

\begin{lstlisting}
python normalize.py --rootdir=/home/map/data/asas11 --dict=dict.sqlite --rawlc=rlc.sqlite rawlcsel="select * from stars where staruid = ? and (quality is NULL or quality = 'A' or quality = 'B') order by hjd asc;" --nplc=nplc.sqlite
\end{lstlisting}

All database files are located in \texttt{rootdir}, we are using \texttt{dict.sqlite} as dictionary, the raw light curves are in \texttt{rlc.sqlite}, the normalized light curves will be written to \texttt{nplc.sqlite} (for normalized, phased light curve). The \texttt{rawlcsel} option contains the SQL select statement responsible for the quality selection. Here we only want measurements with a quality flag of A, B or no quality flag (this is necessary for fields released early by the ASAS team). We want the entries being sorted ascending by Julian date (hjd). The program is running over all stars in the dictionary - the not listed \texttt{select} option for the dictionary defaults to \texttt{select * from stars}, the SQL equivalent of "for every row select all columns from the table stars". For every star the program will delete any previous normalized light curve stored in \texttt{nplc.sqlite}. This can be suppressed using the not listed \texttt{nodel} option and saves a few seconds in run-time when creating the database the first time. However, there should be only one light curve for any given star, so by default previous light curves should be deleted.

The next step is to phase the light curves. There are not many options to set, we just hand over the root directory, the dictionary file and the database for the normalized and phased light curve. Please note, that we do need a period and a zero point offset ($T0$) which we get from the dictionary. We can re-run the program using improved periods or changed offsets for all or individual stars by using the \texttt{select} option, for example \texttt{"select * from stars where id = '000006+2553.2';"} if we want to run the program for the star with the ASAS ID 000006+2553.2. For now, we let it run over all stars:

\begin{lstlisting}
python makeplc --rootdir=/home/map/data/asas --dict=dict.sqlite --nplc=nplc.sqlite
\end{lstlisting}

The third step is to bin the phased light curves in order to prepare them for polyfit. This step is not strictly necessary, but it drastically reduces the failure rate and run-time of polyfit: 

\begin{lstlisting}
python makeblc --rootdir=/home/map/data/asas --dict=dict.sqlite --plc=nplc.sqlite --blc=blc.sqlite --nrbins=50
\end{lstlisting}

The dictionary and the database with the normalized, phased light curves are the input files, the option \texttt{blc} gives the output file for the binned light curves. In this case we choose a maximum of $50$ bins per light curve. Please note, that the bins are equidistant over the phase interval $[-0.5; 0.5[$. If we have gaps due to missing measurements, these gaps will show in the binned light curve as well. 

The fourth and last step of light curve preparation is to get a polynomial fit and plot the phased light curve with its fit. While \texttt{polyfit} itself is written in C, we do not call it directly, but by using a multi-process Python wrapper which calls polyfit for several light curves in parallel. How many parallel processes are started is controlled by the environment variable \texttt{OMP\_NUM\_THREADS} - which is generally used by the OpenMP extension of the gcc-compiler and many other programs. Our command line is:

\begin{lstlisting}
python runpolymp.py --rootdir=/home/map/data/asas --dict=dict.sqlite --plc=nplc.sqlite --blc=blc.sqlite --fit=fit.sqlite --logfile=fiteqlog.txt --select="select * from stars where sdir = 'vareq';"
\end{lstlisting}

The dictionary, phased and binned light curve are the inputs - we need the phased light curves only for the plots. The output is given in the \texttt{fit} option. This time we restrict the number of light curves to those being in the equatorial stripe (having the field \texttt{sdir} set to "vareq"). Please note, that we need to use double quotes here, because \texttt{sdir} is a string variable and we restrict it to have a specific value for which the SQL syntax requires single quotes. Aside of the database with the polynomial fit we will find the sub-directories \texttt{vareq/plots} and \texttt{vareq/failed} within the root directory, they are created if necessary. They hold plots of successfully fitted and failed light curves, example plots are in Figure \ref{lcsamples}. From \texttt{polyfit} we get a reduced $\chi^2$ value as quality of the fit and we store it in the dictionary database. This value is useful for selecting the quality of fitted light curves, we actually train on in the next step.

The fit parameters serve as input either for training a network or for running an already trained network and get a classification. We start with the training. 

\begin{lstlisting}
python trainnetmp --rootdir=/home/map/data/asas  --dict=dict.sqlite --fit=fit.sqlite --minhidden=15 --maxhidden=20 --nrnets=10 --eta=0.60 --resdir=testrun --logfile=testlog.txt --repname=report.txt --classes=classes.txt --clscol=varcls --selfile=select.txt --pname=best.pickle
\end{lstlisting}

Aside of the meanwhile familiar options for the root directory, dictionary and fit database, we train networks with 15 - 20 neurons in the hidden layer, 10 networks for each number of hidden neurons. We use a learning rate \texttt{eta} of $0.60$ and give a directory in \texttt{resdir} where the results, reports and logfiles are stored. The text file given as \texttt{classes} parameter holds a white-space separated line of the class-names we want to train on. This allows to train on less classes than the database actually provides.  The reason is that we want to skip over classes that have too few exemplars to train on (\texttt{PULS} with just seven exemplars for example). By omitting \texttt{PULS} the program will then skip over all stars with this classification. 

We do not give the select statement directly this time, but provide a text file as \texttt{sefile} option that holds a select statement. ASAS provides mixed classes like \texttt{EC/DSC} for a mixture of contact binary and $\delta$ Scuti. Given the fact that ASAS does not provide any probabilities, we do not want to train on mixed classes, so we have to filter them out. This makes the select statement quite long and it is easier to compose, test and store it in a text-file. It looks like this:

\begin{verbatim}
select * from stars 
where varcls not like '%/%' and varcls not like '%=%' and
      varcls not like '%:%' and
      chi2 is not null and chi2 <= 1.0 and
      T0 > 0;
\end{verbatim}

We sort out multiple classes (marked by "/" or "\%" as separator) and any light curve with weak characteristics (marked by ":"). In addition we require the $\chi^2$ value to be set and less or equal 1.0 (high quality fits). Last we require a non-zero offset ($T0$). 

\texttt{trainnetmp} will collect the report for every trained network - training and validation errors, confusion matrix for the testing data set. Based on the overall retrieval rate the best network will be stored as "pickle" file (\texttt{pname} option), one of the methods in Python to store objects. We choose pickle files as output, because we did not want to get into the trouble of conflicting database access if several training runs finish at about the same time. In addition local storage on the ACCRE cluster is much faster than database access over the network. We inspect the reports of several runs and then decide which network will be stored in the database. The conversion is done by \texttt{pickle2db}:

\begin{lstlisting}
python pickle2db.py --rootdir=/home/map/data/asas --resdir=testrun --pname=best.pickle --name=testrun --dbname=net.sqlite
\end{lstlisting}

Aside of the root directory we give the directory name with the results and the name of the pickle file. We name this run in order to distinguish it from other training results and finally give the name for the network database.

Running the network over unclassified light curves is done by \texttt{runtrained}. We can either use a pickle file or our database with trained networks as input:

\begin{lstlisting}
python runtrained.py --rootdir=/home/map/data/externaldb --dict=externaldict.sqlite --fit=externalfit.sqlite --fqnetdb=/home/map/data/asas/net.sqlite --netname=testrun --repname=report.txt --clsname=classification.sqlite --clscol=externalclass
\end{lstlisting}
or 
\begin{lstlisting}
python runtrained.py --rootdir=/home/map/data/externaldb --dict=externaldict.sqlite --fit=externalfit.sqlite --fqpickle=/home/map/data/asas/testrun/best.pickle --repname=report.txt --clsname=classification.sqlite --clscol=externalclass
\end{lstlisting}

Please note, that the root directory, dictionary and fit database are labeled as "external" in order to indicate, that we can classify any other database than ASAS. We refer the trained network either with the fully qualified path to the database and the name we gave it (\texttt{fqnetdb} and \texttt{netname}) or by the fully qualified path to the pickle file \texttt{fqpickle} - if we want a quick test before we store it in the database.

\subsection{Helper Classes and Programs}

In this subsection we briefly mention some programs and classes that might be found useful.

We found that quite some ASAS light curves are not phased correctly and implemented an algorithm based on the minimum, median and maximum value in order to decide if phase 0 should be the minimum or maximum of the light curve (\texttt{rephase}). The algorithm is described in \citet{keplerlle}. The script allows automatic or manual phasing of one or many light curves.

While training a network, the validation error is used to find out when the network has learned all it can. \texttt{qplotlog} allows to drag and drop logfiles written during a training session onto its window and generates a plot. 

\texttt{lcview} allows to query the database for matching or mismatching classifications and to look at one or more light curves in order to decide which  classification is correct. This program is still under heavy construction.

The neural network itself is encapsulated in a Python class named \texttt{mlp}. Although instances of this class are not called from the command line, but are configured and provided with datasets via \texttt{trainnetmp} or \texttt{runtrained}, we want to at least mention the "work horse" doing all the training, validating and testing. 

\clearpage

\section{\label{appendix} Confusion Matrices}

Here we provide the detailed confusion matrices for each of the samples studied. Each table shows the distribution of correctly and incorrectly classified light curves. Table \ref{singleclasscm} gives the confusion matrix for the case of the 'single class' dataset (see Section \ref{sectsingleclass}). Table \ref{cleancm} gives the confusion matrix for the case of 'clean dataset' (see Section \ref{sectcleanlc}). Table \ref{singlewithcleancm} gives the confusion matrix for the case of the 'single class' dataset classified with the classifier trained on the 'clean' dataset. Finally, Table 6 \ref{keplercm} gives the confusion matrix for the case of the Kepler dataset (see \ref{sectsingleclass}).

\begin{table}[h]
    \small
    \begin{tabular}{|l|c|c|c|c|c|c|c|c|c|c|c|}
        \hline
output/target & DC-FO & DC-FU &  DSCT &    EC &    ED &     ESD &    MIRA &    MISC &    RRAB &     RRC &   \% F Pos \\
\hline
DC-FO &    19 &     1 &       0 &     0 &     1 &       1 &       0 &       4 &       2 &       0 &   32.14  \\
DC-FU &     6 &   101 &       0 &     0 &     3 &       1 &       0 &      18 &       0 &       0 &   21.71  \\
 DSCT &     0 &     0 &     130 &     0 &     0 &       1 &       0 &      18 &       0 &       0 &   12.75  \\
   EC &     1 &     0 &       1 &   606 &     4 &      36 &       0 &       7 &       3 &       3 &    8.32  \\
   ED &     0 &     2 &       0 &     4 &   273 &      17 &       0 &       4 &       0 &       2 &    9.60  \\
  ESD &     0 &     1 &       0 &    15 &    14 &     160 &       0 &       1 &       0 &       1 &   16.67  \\
 MIRA &     0 &     0 &       0 &     0 &     0 &       0 &     323 &      47 &       0 &       0 &   12.70  \\
 MISC &    18 &    43 &       1 &     4 &     5 &       6 &      51 &    6006 &       1 &       2 &    2.13  \\
 RRAB &     8 &     0 &       0 &     5 &     2 &       0 &       0 &       5 &     275 &       5 &    8.33  \\
  RRC &     0 &     0 &       1 &     3 &     0 &       4 &       0 &       5 &       1 &      73 &   16.09  \\
\hline
  Sum &    52 &   148 &     133 &   637 &   302 &     226 &     374 &    6115 &     282 &      86 &          \\
\% corr & 36.54 & 68.24 &   97.74 & 95.13 & 90.40 & 70.80 &   86.36 &   98.22 &   97.52 &   84.88 &          \\
\% EBs  &       &       &         & 98.11 & 96.35 & 94.24 &         &         &         &         &          \\
\hline
    \end{tabular}
    \centering
    \parbox{16cm}{
        \caption[Confusion matrix for single classes]{\label{singleclasscm} Confusion matrix for the 'single class' dataset. Target classes as columns, network classification (output) as rows. The last row contains the retrieval rate if eclipsing binaries are seen as one class. 19 neurons in hidden layer, 16708 stars used for training, 8354 stars for validation and 8355 for testing (results in this table) }}
\end{table}
\begin{table}[h]
    \small
    \begin{tabular}{|l|c|c|c|c|c|c|c|c|c|c|c|}
        \hline
output/target & DC-FO & DC-FU &    DSCT &      EC &      ED &     ESD &    MIRA &    MISC &    RRAB &     RRC &   \% F Pos \\
\hline
DC-FO   &    20 &     2 &       0 &       0 &       1 &       0 &       0 &       8 &       0 &       0 &   35.48  \\
DC-FU   &     3 &    67 &       0 &       1 &       1 &       0 &       0 &      11 &       0 &       0 &   19.28  \\
 DSCT   &     0 &     0 &      78 &       0 &       0 &       0 &       0 &       3 &       0 &       0 &    3.70  \\
   EC   &     0 &     0 &       0 &     510 &       0 &      30 &       0 &       4 &       2 &       3 &    7.10  \\
   ED   &     0 &     0 &       0 &       0 &     229 &       6 &       0 &       2 &       1 &       0 &    3.78  \\
  ESD   &     0 &     1 &       0 &       6 &      17 &     142 &       0 &       2 &       0 &       0 &   15.48  \\
 MIRA   &     0 &     0 &       0 &       0 &       0 &       0 &      37 &       2 &       0 &       0 &    5.13  \\
 MISC   &     2 &    27 &       0 &       4 &       3 &       2 &       7 &     834 &       2 &       0 &    5.33  \\
 RRAB   &     7 &     0 &       0 &       2 &       0 &       0 &       0 &       1 &     101 &       0 &    9.01  \\
  RRC   &     0 &     0 &       2 &       0 &       0 &       2 &       0 &       0 &       2 &      42 &   12.50  \\
\hline
  Sum   &    32 &    97 &      80 &     523 &     251 &     182 &      44 &     867 &     108 &      45 &  \\
\% corr & 62.50 & 69.07 &   97.50 &   97.51 &   91.24 &   78.02 &   84.09 &   96.19 &   93.52 &   93.33 &  \\
\% EBs  &       &       &         &   98.66 &   98.01 &   97.80 &         &         &         &         &  \\
\hline
    \end{tabular}
    \centering
    \parbox{16cm}{
        \caption[Confusion matrix for clean light curves]{\label{cleancm} Confusion matrix for the 'clean' dataset (single class in ASAS and $\chi^2 \le 1.0$. Target classes as columns, network classification (output) as rows. The last row contains the retrieval rate if eclipsing binaries are seen as one class. 18 neurons in hidden layer, 4457 stars used for training, 2228 stars for validation and 2229 for testing (results in this table) }}
\end{table}

\begin{table}[h]
    \small
    \begin{tabular}{|l|c|c|c|c|c|c|c|c|c|c|c|}
        \hline
output/target & DC-FO & DC-FU &    DSCT &      EC &      ED &     ESD &    MIRA &    MISC &    RRAB &     RRC &   \% F Pos \\
\hline
DC-FO   &    70 &    19 &       0 &       7 &       5 &       4 &       0 &      72 &       5 &       0 &   61.54 \\
DC-FU   &    29 &   335 &       0 &       7 &      12 &       6 &       1 &     285 &       0 &       0 &   50.37 \\
  SCT   &     0 &     0 &     514 &       4 &       0 &       4 &       0 &      76 &       2 &      22 &   17.36 \\
   EC   &     5 &     7 &       3 &    2449 &      11 &     158 &       4 &     133 &     102 &      27 &   15.52 \\
   ED   &    12 &     7 &       0 &      11 &    1010 &      51 &       0 &      40 &      11 &       3 &   11.79 \\
  ESD   &     9 &     8 &       0 &     104 &      76 &     576 &       0 &      81 &      26 &       3 &   34.77 \\
 MIRA   &     0 &     4 &       0 &       1 &       0 &       0 &    1292 &     305 &       0 &       0 &   19.35 \\
 MISC   &    21 &   194 &      13 &      30 &      34 &      32 &     295 &   23259 &       9 &       7 &    2.66 \\
 RRAB   &    35 &     5 &       0 &      33 &      10 &       7 &       0 &      35 &    1058 &      13 &   11.54 \\
  RRC   &     0 &     0 &      12 &      15 &       1 &      11 &       0 &      15 &      12 &     253 &   20.69 \\
\hline
  Sum   &   181 &   579 &     542 &    2661 &    1159 &     849 &    1592 &   24301 &    1225 &     328 &         \\
\% corr & 38.67 & 57.86 &   94.83 &   92.03 &   87.14 &   67.84 &   81.16 &   95.71 &   86.37 &   77.13 &         \\
\% EBs  &       &       &         &   96.35 &   94.65 &   92.46 &         &         &         &         &         \\
\hline
    \end{tabular}
    \centering
    \parbox{16cm}{
        \caption[Confusion matrix for the single class set run by a net trained with clean light curves]{\label{singlewithcleancm} Confusion matrix for the single class dataset by a network trained with clean light curves. Target classes as columns, network classification (output) as rows. The last row contains the retrieval rate if eclipsing binaries are seen as one class.}}
\end{table}

\begin{table}[h]
    \small
    \begin{tabular}{|l|c|c|c|c|}
        \hline
output/target &       EC   &     ED   &     ESD   &    \% F Pos  \\
\hline
DC-FO   &        0   &      3   &      0    &      \\
DC-FU   &        0   &     20   &      0    &      \\
 DSCT   &        1   &      0   &      0    &      \\
   EC   &      374   &     31   &     47    &    17.26   \\
   ED   &        0   &    427   &     13    &     2.95   \\
  ESD   &       64   &    134   &     68    &    74.44   \\
 MIRA   &        3   &      0   &      1    &      \\
 MISC   &        0   &    453   &      6    &      \\
 RRAB   &        1   &     19   &      2    &      \\
  RRC   &        2   &      6   &      4    &      \\
\hline
  Sum   &      445   &   1093   &    141    &      \\
\% corr &     84.04  &   39.07  &   48.23   &      \\
\% EBs  &     98.65  &   54.16  &   90.78   &      \\
\hline
    \end{tabular}
    \centering
    \parbox{16cm}{
        \caption[Confusion Matrix for the \textit{Kepler} Data Set by a Network Trained with Single Class Light Curves]{\label{keplercm} Confusion matrix for the \textit{Kepler} Data Set by a Network Trained with Single Class Light Curves. Target classes as columns, network classification (output) as rows. The last row contains the retrieval rate if eclipsing binaries are seen as one class.}}
\end{table}

\end{appendices}

\clearpage
\bibliography{myrefs}

\clearpage
\end{document}